\documentclass[12pt]{emulateapj}

\shorttitle{Stellar Mass Tully-Fisher to $z=1.2$}
\shortauthors{Kassin et al.}

\slugcomment{Accepted for Publication in the ApJL}

\begin{document}

\title{The Stellar Mass Tully-Fisher Relation to $z=1.2$ from AEGIS}

\author{Susan A. Kassin\altaffilmark{1},
Benjamin J. Weiner\altaffilmark{2}, S. M. Faber\altaffilmark{1}, David C. Koo\altaffilmark{1}, Jennifer M. Lotz\altaffilmark{3}, J$\rm \ddot{u}$rg Diemand\altaffilmark{1,}\altaffilmark{4}, Justin J. Harker\altaffilmark{1}, Kevin Bundy\altaffilmark{5}, A. J. Metevier\altaffilmark{1}, Andrew C. Phillips\altaffilmark{1},  Michael C. Cooper\altaffilmark{6}, Darren J. Croton\altaffilmark{5}, Nicholas Konidaris\altaffilmark{1}, Kai G. Noeske\altaffilmark{1}, \& C. N. A. Willmer\altaffilmark{2,}\altaffilmark{7}}

\altaffiltext{1}{UCO/Lick Observatory, Department of Astronomy \& Astrophysics, University of California, Santa Cruz, CA 95064; kassin, faber, koo, jharker, anne, phillips, npk, kai, and diemand@ucolick.org}
\altaffiltext{2}{Steward Observatory, University of Arizona, Tucson, AZ; bjw@as.arizona.edu, cnaw@as.arizona.edu}
\altaffiltext{3}{Leo Goldberg Fellow, National Optical Astronomical Observatories, Tuscon, AZ; lotz@noao.edu}
\altaffiltext{4}{Hubble Fellow}
\altaffiltext{5}{Reinhardt Fellow, University of Toronto, Toronto, Canada; bundy@astro.utoronto.ca}
\altaffiltext{6}{Department of Astronomy, University of California, Berkeley, CA; cooper@astron.berkeley.edu, darren@astron.berkeley.edu}
\altaffiltext{7}{On leave from Observat\'orio Nacional, Brazil.}

\begin{abstract}
We combine newly measured rotation velocities, velocity dispersions, and
stellar masses to construct stellar mass Tully-Fisher relations
(M$_*$TFRs) for 544 galaxies with strong emission lines at $0.1<z<1.2$
from the All Wavelength Extended Groth Strip International Survey (AEGIS) and the
Deep Extragalactic Evolutionary Probe 2 Survey (DEEP2).  The conventional M$_*$TFR using 
only rotation velocity ($V_{rot}$) shows large scatter ($\sim 1.5$ dex in velocity).  The
scatter and residuals are correlated with morphology in the sense that disturbed, compact,
and major merger galaxies have lower velocities for their masses.
We construct an M$_*$TFR using the kinematic estimator $S_{0.5}$
which is defined as {\small $\sqrt{0.5V_{rot}^2 + \sigma_g^2}$} and accounts for disordered or 
non-circular motions through the gas velocity dispersion ($\sigma_g$).   The new M$_*$TFR,
termed $S_{0.5}$/M$_*$TFR, is remarkably tight over $0.1<z<1.2$
with no detectable evolution of its intercept or slope with redshift.
The average best fit relation has 0.47 dex scatter in stellar mass, 
corresponding to $\sim 1.2$ `magnitudes,' assuming a constant mass-to-light ratio.
Interestingly, the $S_{0.5}$/M$_*$TFR  is consistent with
the absorption-line based stellar mass Faber-Jackson relation for nearby elliptical galaxies
in terms of slope and intercept, which might suggest a physical connection between the two relations.
\end{abstract}


\section{Introduction}
The Tully-Fisher relation (TFR) between the luminosity of galaxies and 
their rotation velocity is a fundamental scaling relation that constrains 
galaxy formation and evolution models \citep[e.g.,][and references therein]{dalc}. Recent local studies 
focus on stellar or baryonic masses \citep[e.g.,][]{bdej,mcg5,piza}, which are easier to model than luminosity.  Distant stellar-mass TFRs (M$_*$TFRs) have also been studied (Conselice et al. 
2005; Flores et al. 2006). Neither found evidence for evolution to 
redshifts $z \sim 1$ or 0.6, respectively, but each pruned their data to 
exclude morphologically odd galaxies. This letter presents a study of the 
M$_*$TFR to redshift $z = 1.2$ using a sample of 544 galaxies with strong 
emission lines.  Following \citet{wei2}, we exclude few 
galaxies and adopt a new velocity measure that includes both velocity 
gradients (usually rotation) and velocity spread (dispersion). We adopt a 
$\Lambda$CDM cosmology with $h=0.7$, $\Omega_m=0.3$, and $\Omega_{\Lambda}=0.7$.

\section{Observations}
Galaxy kinematics and  morphologies  at $z \sim 1$ are possible with high 
resolution spectra and images. To track different morphologies and 
redshifts, samples should  be large enough to be subdivided. The
All-Wavelength Extended Groth Strip International Survey (AEGIS)
provides such data.  Single-orbit images were taken by the 
Hubble Space Telescope Advanced Camera for Surveys (HST/ACS) in the F606W ($V$) 
and F814W ($I$) band-passes, and spectra were obtained by the
Deep Extragalactic Evolutionary Probe 2 Survey (DEEP2); see details in \citet{davi,lead}.
Observations and reduction of kinematics are described in \citet{wei1}
for data similar to DEEP2.
Kinematics are measured from the strongest emission line(s) among H$\alpha$, H$\beta$,
[OII] $\lambda 3727$, and [O III] $\lambda 5007$.
The spectral resolution is FWHM=$1.4$ \AA, or 56 km $s^{-1}$ at $z=1$ 
(Gaussian $\sigma=24$ km s$^{-1}$). 
Emission line widths can be measured down to $\sim 0.6$ of the
$\sigma$ or $\sim 15$ km s$^{-1}$, and velocity centroids for rotation curves
are good to 0.4 of $\sigma$ or $\sim 10$ km s$^{-1}$.  Rotation velocities can be
detected down to $\sim 5$ km s$^{-1}$, if present. 

\section{Sample Selection}
We chose a sample of galaxies that span a wide range in redshift, stellar 
mass, and morphology.
The DEEP2 survey had magnitude limits of $R_{AB}=24.1$, 
bright enough to yield good HST/ACS imaging.    A redshift limit of $z=1.2$ ensures
that galaxies of all colors on the ``blue sequence" \citep{will} are well-sampled.
To yield good quality kinematics, emission lines were required to have integrated intensities 
$> 1500$ e$^{-}/$\AA\ in the summed one-dimensional spectrum,
and spectrographic slits had to be aligned to within 40\degr\ of the 
major axes of the galaxies (see Fig. 13 of \citealt{wei1}). Also, to reduce the 
effects of dust, edge-on systems with $i > 70$ were excluded, and to allow a 
reliable rotation measure, nearly face on galaxies with $i < 30$ were 
removed. A total of 14 galaxies were excluded because they had emission lines with disturbed morphologies or that had contamination by emission from a nearby galaxy on the sky.
After additional cosmetic cuts were made \citep{kass}, 
the final sample became 544 galaxies.  Our final sample is primarily selected on emission line strength.

\section{Data Reduction and Analysis}
\subsection{Photometric Parameters}
In order to interpret dynamical measurements made from the spectra, 
accurate position angles and ellipticities of galaxies must be obtained from HST
imaging. To measure these parameters, 
the SExtractor galaxy photometry software \citep{bert} is used, as discussed in \citet{lotz}.
To obtain quantitative morphologies for galaxies, we use the Gini/M$_{20}$ method of \citet{lot1}.
This system of objective structural parameters has been shown to reliably
divide galaxies into bins analogous to Hubble types; namely, early type spirals, late type 
spirals and irregulars, and major mergers.
Galaxies are classified in the $V$ image for $z<0.6$ and in $I$
for $z\geq 0.6$.  Such quantitative morphologies are obtainable only for galaxies with Petrosian 
radii $> 0.3 \arcsec$, images that have a signal-to-noise ratio of $> 2.5$, 
and that do not fall near the edge of the HST/ACS CCD chip.
Galaxies without quantitative morphologies are not removed from the sample
for completeness sake, and do not differ statistically from the rest
of the sample.  All photometric parameters, 
including quantitative morphologies, will appear
in a forthcoming paper, \citet{lot2}.  In a separate exercise, we visually classified 
all galaxies and found examples that were fairly normal according to Gini/M$_{20}$, 
but looked to our eye to be more disturbed or compact; we discuss these galaxies
separately.

\subsection{Spectroscopic Measurements}
Spatially extended emission lines in the galaxy spectra are used to measure 
gas rotation and dispersion profiles. 
Since the spatial extent of the line emission is at most only a few times the
seeing (typically $0.7\arcsec$ FWHM), the effect of seeing must be modeled.  
We do this with the ROTCURVE fitting procedure of \citet{wei1}.  The kinematic
model we fit has two parameters: the velocity on the flat part of the rotation 
velocity profile, $V_{rot}$, and the velocity dispersion from a spatially resolved 
fit to the two-dimensional spectra, $\sigma_g$.  The best-fit $V_{rot}$ 
values are corrected for inclination with the measured galaxy ellipticities.
For galaxies without a disk-like geometry
(based on visual inspection of HST/ACS images), $V_{rot}$
is not inclination-corrected.  These galaxies are flagged
in the following analysis, but fits do not change
if all galaxies are inclination-corrected.  The $\sigma_g$ we measure does 
not have to correspond to a literal gas velocity dispersion like that of stars in an elliptical galaxy.
Instead, for $\sigma_g \la 20$ km s$^{-1}$, $\sigma_g$ likely measures the relative 
motions of individual \ion{H}{2} regions, as in spiral arms or a thick disk;
for $\sigma_g \ga 20$ km s$^{-1}$,
$\sigma_g$ likely represents an effective velocity dispersion caused by the blurring of velocity
gradients on scales at or below the seeing limit that may not even have a preferred plane.
An uncertainty of 0.1 dex is adopted in both $\sigma_g$ and the inclination-corrected
$V_{rot}$ to account for random errors and the dependence of the 
model parameters on the assumed seeing and scale radius of the
rotation curve.  Results do not change significantly if uncertainties of 0.2
are chosen, as discussed in \S 6. 

\section{Stellar Mass Estimates}
A few methods of estimating the stellar masses ($M_*$) of galaxies have been investigated
in a separate study \citep{kass}.
These methods are: a color-$M/L$ relation from \citet{bell}
for $B-V$ and $M/L_B$, a color-$M/L$ relation from \citet{bell} for $B-V$ and $M/L_H$, 
$M/L_H=1$, broad-band spectral energy distribution (SED)
fits of \citet{bund} which incorporate observed optical and $K$-band data, 
and the method of \citet{lihw}.
We choose to adopt the method of \citet{lihw}.
These authors calibrated rest-frame $UBV$ photometry and redshifts to the full SED fits
of Bundy et al.\ (2005) to obtain $M_*$ estimates for very blue 
galaxies that are not detected at $K$, which Bundy et al.\ (2005) require
to derive a $M_*$.  While the evolution of SEDs with redshift is included and 
errors of 0.2 dex are assumed, the results of this paper, as discussed next,
change insignificantly if we do not take into account evolution or increase the errors to 0.3 dex
\citep{kg07}.

\section{$S_{0.5}$ as a Tracer of Galaxy-Dark Halo Potential Wells}
The quantity $S_K^2 \equiv KV_{rot}^2 + \sigma_g^2$, where $K$ is a constant,
combines dynamical support from ordered motion 
with that from disordered motions \citep{wei1}.  For a spherically symmetric tracer population
with isotropic velocity dispersion and density 
$\propto r^{-\alpha}$, $\sigma_g=V_{rot}/\sqrt{\alpha}$ \citep[][$\S 4.4$]{binn}, where $K\equiv 1/\alpha$.
Therefore, if galaxies are virialized systems, and these assumptions about the tracer population 
are at least approximately correct, then $S_K$ should at least approximately trace the global $\sigma$ of galaxy-halo systems.  In this case, the tracer population is the gas producing
the emission lines, which we assume follows the $M_*$.  For disk galaxies with
exponential $M_*$ profiles, $\alpha \simeq 2-3$ brackets a range of reasonable values 
(\S5.2  of \citealt{wei1}).
A satisfactory approximation is adopted for all morphologies,
$K=\onehalf$;  if $K=\onethird$ is adopted, the overall results do not change.
We henceforth use $S_{0.5}$ for our analysis. 

\section{Results}
The top panel  of Figure 1 shows $V_{rot}$ versus $M_*$ (the $V_{rot}/M_*$TFR) for a range of
redshifts.   The majority of spirals (green triangles and blue squares not outlined in black)
form a clear ridge-line that compares well to a local relation from \citet{bdej}, consisting of well-ordered
spirals, and to a $z\sim1$ relation from \citet{cons}, who included only ``elongated disks."
In contrast, almost all of the other classes (red circles and symbols outlined in black) 
lie to lower $V_{rot}$, causing a large scatter ($\sim 1.5$ dex in velocity).

\begin{figure*}[th!]
\epsscale{.55}
\vspace{0.3in}
\includegraphics[scale=0.85]{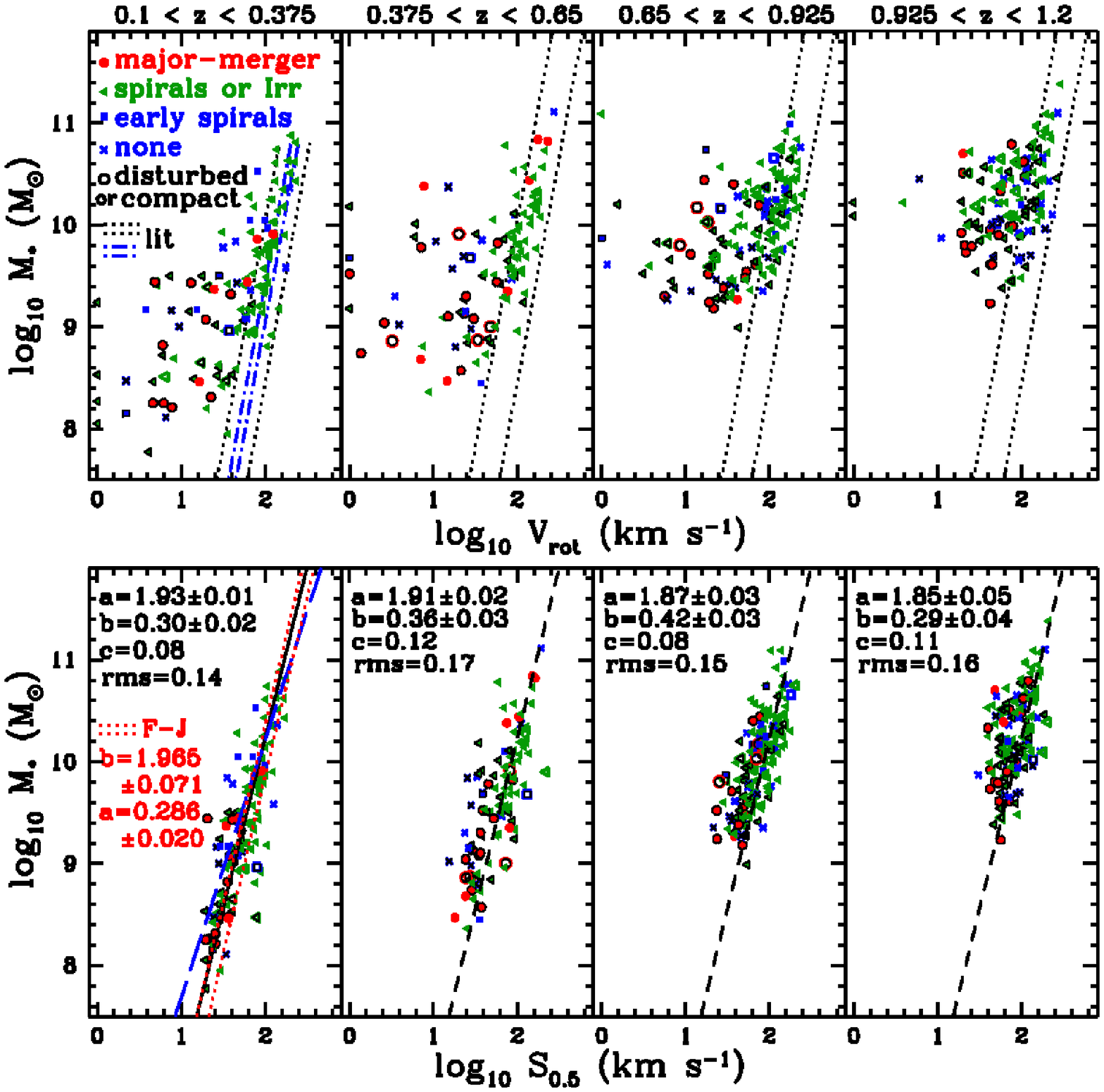}
\vspace{0.4in}
\caption{\normalsize The $S_{0.5}$ and $V_{rot}$ stellar-mass Tully-Fisher 
relations (M$_*$TFRs) for $0.1<z<1.2$ in bins of $z$.  
Galaxies are plotted as different symbols according to their quantitative morphologies,
and if a morphology is unobtainable, it is placed into the `none' category.
Symbols are outlined in black if the galaxies 
were considered to be disturbed or compact via a separate visual classification. 
If $V_{rot}$ is not inclination corrected, open symbols are used; all galaxies in the `none'
category are inclination-corrected as their symbol does not have an open form.
Using $S_{0.5}$, which combines the dynamical support from ordered motion with
that from disordered motions, results in an M$_*$TFR that is much tighter
and non-evolving to $z=1.2$.
For $S_{0.5}$, a fit to data in the lowest $z$ bin is plotted as a solid line
that is repeated in the other $z$ bins as dashed lines.  
Fits are labeled in each box, where {\it a} is the intercept at $10^{10} \rm M_{\odot}$,
{\it b} the slope, {\it c} the intrinsic scatter in $S_{0.5}$, and {\it rms} the
rms scatter in $S_{0.5}$.  In the lowest $z$ bin for $V_{rot}$, the M$_*$TFRs from local and
$z\sim 1$ studies (\citealt{bdej} and \citealt{cons}, respectively) are plotted as two dot-dashed and
dotted lines, respectively, to delineate rms scatter. 
In the lowest $z$ bin for $S_{0.5}$, the local $M_*$ Faber-Jackson relation 
\citep{gall} is plotted as two dotted lines to delineate rms scatter.  All relations from the literature have been converted to 
a Chabrier initial mass function when necessary.  An estimate of the $S_{0.5}/M_{baryon}$TFR
(\S 9) is plotted as a long dashed blue line in the lowest $z$ bin.}
\end{figure*}

This morphological dependence of Tully-Fisher scatter is well known.  Among local
studies, e.g., larger scatter is found for close pairs with kinematic disorders \citep{bart}
and when peculiar galaxies are included \citep{kann}.  At higher redshifts, larger scatter is 
found for $z\sim0.5$ galaxies with ``complex kinematics" \citep{flor}.  Moreover, 
$\sim 25$\% of galaxies found with kinematics unrelated to rotation are excluded from
Tully-Fisher studies at $0.25<z<0.45$ \citep{sp98} and at $0.1<z<1.0$  \citep{bohm}.
In summary, because of morphological pruning, no study except that of \citet{wei2}
incorporates the population of Tully-Fisher outliers that we detect in a measurement of Tully-Fisher
evolution, and the conclusions of such studies will only apply for
the subsample of galaxies that are morphologically well-ordered.

For $M_*>10^{10}M_{\odot}$, the percentage of galaxies that scatter to low $V_{rot}$
from the Tully-Fisher ridge-line (defined as the left-most dotted line in Figure 1)
increases with redshift (18\%, 35\%, 42\%, and 62\% 
for the four redshift bins plotted).  Similarly, \citet{wei1} find
bright galaxies with $\sigma_g > V_{rot}$ at $z \ge 0.5$ that are rare at low redshift.

The bottom panel of Fig.\,1 uses $S_{0.5}$ instead of $V_{rot}$
in the M$_*$TFR.  The resulting relation has scatter that is not much  
greater than our measurement uncertainties and
does not evolve significantly with $z$.\footnote{One would expect a small 
($\sim 0.05$ dex) decrease in intercept from $z=1$ to 
$z=0$ due simply to the predicted increase in $M_*$ of $\sim 50\%$  for a typical 
spiral galaxy ($z_{formation}=5$ and an exponentially decreasing
star formation rate with $\tau=7$ Gyr), but it would not be detectable given our 
errors.  We do not attempt to determine $K$ by minimizing the scatter in the
$S_K/M_*$TFR because $K$ is sensitive to the morphological mix of the
sample and the distribution of measurement errors.}  Evidently, galaxies with low $V_{rot}$ 
compared to the $V_{rot}$/M$_*$TFR ridge-line have a significant $\sigma_g$ component 
which causes them to lie on the $S_{0.5}$/M$_*$TFR.  
Lines are fit to the $S_{0.5}$/M$_*$TFR in the four redshift bins in Fig.\,1 with a maximum likelihood
method detailed in \citet{wei2}.
We fit log$_{10}\,S_{0.5}= a + b\, \rm log_{10}\, \it M_*$ and all excess scatter is 
allocated to the $S_{0.5}$-coordinate.  The fits do not vary significantly
for the different redshift bins.  Average values
for the slope, intercept at $10^{10} M_{\odot}$, and intrinsic scatter over all 
redshifts are $0.34 \pm 0.05, 1.89 \pm 0.03$, and 0.10 dex in $S_{0.5}$. 
The average rms scatter is 0.16 dex in $S_{0.5}$, which
corresponds to 0.47 dex in $M_*$ ($\sim1.2$ magnitudes). 
The fitted relations are not very sensitive to the error estimates
for $S_{0.5}$ and $M_*$; changing either error estimate by $\pm0.1$ dex 
changes fit intercepts within the 1-$\sigma$ error, and slopes
by not more than $\pm 0.1$ dex.  
However, our limited $M_*$ range at
higher redshift makes slope measurements more uncertain than the
formal fit errors.  Furthermore,
different estimates of $M_*$ can yield a range of $\sim 0.2$ dex in intercept, and $M_*$ estimators with non-evolving SEDs can cause the $S_{0.5}/M_*$TFR to show $\sim 0.1$ dex evolution in intercept with redshift, but this reflects problems of these estimators rather than actual evolution in the $S_{0.5}/M_*$TFR.   

Since no Tully-Fisher sample at low or high redshift has as much scatter 
as our $V_{rot}$/M$_*$TFR due to morphological pruning, it is rather remarkable 
that the $S_{0.5}$/M$_*$TFR
scatter for the same sample of galaxies manages to approach as close
as it does to the scatters found for the $V_{rot}$/M$_*$TFRs for 
galaxies with relatively undisturbed morphologies ($\sim 0.6$ dex in $M_*$ for \citealt{cons}).  
In addition, Fig.\,1 shows that the $M_*$ Faber-Jackson relation 
for low redshift ($0.005 < z\leq 0.22$) early type galaxies from \citet{gall}
is consistent with the $S_{0.5}$/M$_*$TFR in terms of slope and intercept. 

\section{Discussion}
If those galaxies with $M_*>10^{10}$ M$_{\odot}$ that are off the $V_{rot}/M_*$TFR ridge-line with low $V_{rot}$ at higher redshift eventually form into well-ordered disk galaxies on the ridge-line, this trend implies that more massive disk galaxies may be moving onto the $V_{rot}$/M$_*$TFR with time as their stars and gas ``settle" into more circular/planar-dominated orbits.
The possible mechanisms behind this settling can be circularization of gas orbits, 
gaseous dissipation to the disk plane, and/or mergers ending with the
growth of a gaseous disk.  The scatter for $M_* \leq 10^{10} \rm M_{\odot}$, 
which we can observe at lower redshift, is also large, suggesting that lower $M_*$ 
galaxies may currently be settling.  

A scenario of galaxy formation that is consistent with our results
is one that begins with matter assembling in 
dark halo potential wells with random orbital kinematics.  The baryonic components
form proto-disks that are initially supported by a combination of
ordered and random motions.  The material in these proto-disks has
been settling since then, unless they undergo major mergers.
Over the last half or so of their lives, proto-disks and their descendants lie on 
nearly the same $S_{0.5}$/M$_*$TFR.  At first glance, the similarity between the
Faber-Jackson relation and the $S_{0.5}$/M$_*$TFR in slope and intercept is reassuring, given
that merging proto-disks on parabolic binary orbits should put the
merger products close to the Faber-Jackson relation in Fig.\,1 \citep[e.g.,][]{robe}. 
This suggests that the origin of the Faber-Jackson relation could in fact be the
pre-existing $S_{0.5}$/M$_*$TFR for disk-like galaxies, if merging is the process that creates
ellipticals.  However, additional studies are needed to confirm this speculative scenario.

\section{Relation to Halo Kinematics}
If the framework we adopt is correct, halo kinematic properties should be related to 
measurable galaxy kinematics. In particular, N-body models show that the slope of the 
relation between halo  mass ($M_s$) and halo $\sigma$ ($\sigma_s$), both measured within the scale radius 
\citep[][]{nava}, is 0.33: log$_{10}\, \sigma_s$ = 0.33 log$_{10}\, M_s$ + constant.  Remarkably, this 
slope is nearly consistent with {\it both} galaxy relations.  
In addition, neither the halo $\sigma_s-M_s$ relation, nor $M_s$ or $\sigma_s$
individually, is predicted to evolve to within 
$\sim10\%$ from $z=1.2$ to now (\citealt{diem}; see \citealt{wech} for evolution in terms of
virial quantities).  So, if $S_{0.5}$ is linearly related to halo $\sigma_s$ and 
$M_*/M_s=$constant, then the $S_{0.5}$/M$_*$TFR should also evolve little from
$z=1.2$, in agreement with observations.  Furthermore, that neither $M_s$ nor $\sigma_s$ 
individually evolves might imply that absolute changes in inner baryon structures are also small.

However, when comparing galaxies to halos, there are at least three major complications.  
First, since $M_*$ is subject to the star-formation history of a galaxy, it
is not as fundamental a quantity as baryonic mass ($M_{baryon}$).  Locally, low-$M_*$ galaxies
have a greater gas-to-star fraction than more massive galaxies
\citep[e.g.,][]{mcg5}, so the low-redshift $S_{0.5}$/M$_{baryon}$TFR ought to have a 
steeper slope than the $S_{0.5}$/M$_*$TFR.  Indeed, if we assume that 
$M_{gas}=0.56M_* + 3.96$ (roughly consistent with data from \citealt{mcg5} using relations
from \citealt{bell} to obtain $M_*$, which we later convert to a \citealt{chab} initial mass function), then the 
$S_{0.5}$/M$_*$TFR slope in the lowest redshift bin steepens to $\sim 0.39$.  Secondly, galaxies should convert gas into stars over time, causing evolution in $M_*/M_s$.  
The third issue is the complicated relation between 
$S_{0.5}$ and $\sigma_s$, which is expected to be affected by baryon dissipation and the response of the dark halo to it \citep[e.g.,][]{blum}, neither of which are completely understood.
With a better understanding or measurement of these factors and their time evolution, the relationship between galaxy and halo scaling laws will ultimately become clearer and the
effects of baryonic processes during galaxy formation may even be elucidated.

\acknowledgments
We acknowledge the following grants: NSF AST-0507483, HST GO-10134, 
and HST AR-09936 and the very significant cultural role and reverence that the 
summit of Mauna Kea has always had within the indigenous Hawaiian community 
and appreciate the opportunity to conduct observations from this mountain.
SAK would like to thank A. Faltenbacher, B. Holden, 
and Z. Zheng for helpful conversations.

\end{document}